\documentclass[reprint,
 amsmath,amssymb,
 aps,
]{revtex4-2}
\usepackage[english]{babel}
\usepackage{csquotes}
\usepackage{graphicx}
\usepackage{dcolumn}% Align table columns on decimal point
\usepackage{bm}% bold math
\usepackage{color}
\newcommand{\change}[1]{#1}
\newcommand{\newchange}[1]{{#1}}

\begin{document}

\preprint{APS/123-QED}

\title{Anisotropic interactions induce \newchange{dynamical arrest} in artificial colloidal ice.}
% \thanks{A footnote to the article title}

\author{Leonardo G. Alanis-Cantú}
\author{Antonio Ortiz-Ambriz}
\email{aortiza@tec.mx}
\affiliation{
Tecnologico de Monterrey, Escuela de Ingeniería y Ciencias\\
Ave. Eugenio Garza Sada 2501, Monterrey, 64849, México.
}

\date{\today}

\begin{abstract}
Artificial Colloidal Ice is an ice-like system used to study the effects of frustration in controlled environments where all degrees of freedom can be accessed \change{at a length-scale large enough for optical visualization and in real time}. We modify this model system by inducing anisotropic interactions through an in-plane magnetic field. \change{In this new regime}, the system has a well-defined ground state consisting of a checkerboard pattern of \change{fully} charged vertices. However, Brownian Dynamics simulations are unable to reach this ground state and instead remain frozen in metastable disordered states, even in the absence of quenched disorder in the lattice. This arrest is caused by the local magnetic enhancement of the potential barrier that the particles need to cross to find a lower energy state. 
\end{abstract}


\maketitle

%\tableofcontents

\section{\label{sec:introduction}Introduction}

The spin glass model was introduced in 1975 \cite{edwards_theory_1975} to explain the unusual behavior of the specific heat in dilute magnetic alloys observed in the late 50s and early 60s \cite{de_nobel_specific_1959,zimmerman_low-temperature_1960}. In this model, the Ising spins interact on a lattice through random coupling energies that produce a corrugated potential landscape in phase space. The quenched disorder produces low-energy states that lack long-range order \change{but have long relaxation times, effectively making them \emph{frozen} in time.} \change{Quenched disorder typically plays a significant role in the freezing mechanism, as has been showed experimentally in materials like $\mathrm{URh}_2 \mathrm{Ge}_2$ where the exhibited spin freezing can be lifted when the source of disorder is removed \cite{sullow_spin_1997}.}
However, macroscopic signatures of spin freezing have also been observed in very pure samples of bulk materials with no disorder, such as $\mathrm{Pr} \mathrm{Au}_2\mathrm{Si}_2$\cite{krimmel_spin-glass_1999}, Gadolinium Gallium Garnet (GGG) \cite{schiffer_frustration_1995}, various magnetic materials \cite{goremychkin_spin-glass_2008,hadouchi_unconventional_2019,zhang_unusual_2008}, and magnetic Pyrochlores \cite{gaulin_spin_1992,raju_magnetic-susceptibility_1992,reimers_short-range_1991,snyder_how_2001,snyder_low-temperature_2004}. The origins of this \emph{unconventional spin freezing} appear to be multiple, from induced dynamical disorder \cite{thygesen_orbital_2017} to lattice distortions induced by fluctuations \cite{mitsumoto_spin-orbital_2020, mitsumoto_replica_2023}.

Artificial Ice systems are engineered to study geometric frustration in a controlled environment. In Artificial Spin Ice (ASI), ferromagnetic elongated nanoislands are fabricated such that they are uniformly magnetized in one direction. These nanoislands can then be arranged in many different spatial configurations to obtain very diverse behaviors \cite{wang_artificial_2006, lao_classical_2018, rougemaille_artificial_2011, mengotti_real-space_2011, andersson_thermally_2016}, including glass transitions \cite{morley_vogel-fulcher-tammann_2017, saccone_direct_2022}, and ergodicity breaking \cite{saccone_real-space_2023}.
Artificial Colloidal Ice (ACI), on the other hand, is composed of interacting colloidal particles confined to move inside bistable potentials \cite{ortiz-ambriz_colloquium_2019}. This system is similar to ASI, but the configuration is easier to observe experimentally as colloidal particles can usually be seen through an optical microscope, and their interactions are tunable \cite{ortiz-ambriz_engineering_2016}. An analogy between both systems can be made by considering the empty space in the bistable potential as a negative topological charge and then defining a dipole for each trap. It was shown in Refs. \cite{nisoli_unexpected_2018} that for infinite, single coordination lattices with isotropically repulsive particles, the topological dipoles in ACI \change{are energetically equivalent to the} magnetic dipoles in ASI. 

Experimentally, ACI is realized by placing superparamagnetic colloids in lithographically etched grooves, \change{and generating an external magnetic field that induces the interactions}. When the magnetic field is perpendicular to the plane where particles sediment, the interactions are \emph{isotropic} and repulsive. Most studies of colloidal ice have maintained this \emph{regime}, either with such a magnetic field \cite{ortiz-ambriz_engineering_2016, oguz_topology_2020, rodriguez-gallo_geometrical_2023} or by considering electrically charged colloids \cite{libal_realizing_2006}. \change{For a review of particle ice systems, the reader is referred to Ref. \cite{ortiz-ambriz_colloquium_2019}.}

\begin{figure}[tb]
    \includegraphics{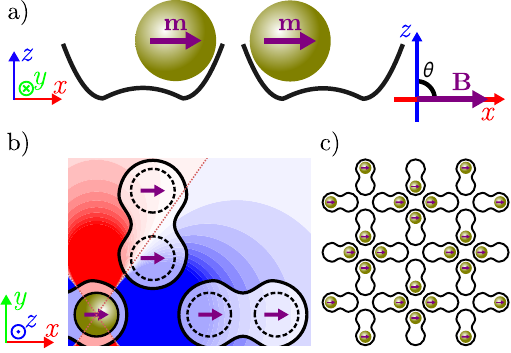}
    \caption{\change{Schematic of the system. a) Colloidal particles are placed in double-well traps with induced magnetic moments pointing along the $x$ axis. b) The in-plane magnetic field produces mixed attractive (blue) and repulsive (red) anisotropic interactions interactions at the single-vertex level. c) The system is arranged in a square lattice such that the magnetic field falls on one of the base vectors. The depicted arrangement corresponds to the ground-state configuration under the studied conditions.}}
\label{fig:system} 
\end{figure}

\change{In this study} we perform simulations that break this condition by applying an in-plane magnetic field, which causes the interaction between colloidal particles to become anisotropic and to have \change{both} attractive and repulsive regions, as shown in Fig. \ref{fig:system}. This causes the system to \newchange{arrest}, even in the absence of any quenched disorder; that is, even if the system has a definite ground state,  \change{it will fall into a metastable state, where due to the local repulsive force, the particles are unable to find a clear path to their lowest energy configuration. The system will then \newchange{stop evolving and stay in this configuration for a very long time}.
We find that these long-lived metastable states appear due to the local increase in the potential barrier produced by the anisotropic interactions, even if the underlying lattice remains completely ordered, which makes this mechanism distinct from the one used to explain spin freezing in Refs. \cite{thygesen_orbital_2017, mitsumoto_spin-orbital_2020}.
}

%%%%%%% Methods

\section{\label{sec:methods}Methods}

For our simulations, we use Brownian Dynamics to compute the position $\mathbf{r}_i$ of particle $i$ at each timestep by integrating the overdamped equations of motion.
\begin{eqnarray}
    \gamma \dfrac{d \mathbf{r}_i}{dt} = \mathbf{F}_i^{\textrm{dd}} + \mathbf{F}_i^{\textrm{trap}} + \bm{\eta},
\end{eqnarray}
where $\gamma = k_BT/D$ is the drag coefficient with diffusion constant $D = 0.14 \mu^2\mathrm{m}/\mathrm{s}$ and $\bm{\eta}$ is a random variable representing the thermal noise, with zero mean $\langle \bm{\eta}\rangle = 0$, and $\langle\bm{\eta}(t)\bm{\eta}(t') \rangle  = \newchange{4}k_BT\gamma\delta(t-t')$; in this case, $T=300$K. 
The term $\mathbf{F}_i^{\textrm{trap}}$ arises from a bistable potential, like the one illustrated in Fig. \ref{fig:system}a) with two components of the force vector: one parallel to the direction of the trap $\hat{\mathbf{e}}_\parallel$ and one perpendicular to the direction of the trap $\hat{\mathbf{e}}_\perp$ which prevents the colloids from jumping to other traps. The full expression is: 
\begin{eqnarray}
    \mathbf{F}_i^{\textrm{trap}} &=& -k_\textrm{trap}r_\perp \hat{\mathbf{e}}_\perp \nonumber \\ 
    &-& \hat{\mathbf{e}}_\parallel 
    \begin{cases}
        -k_{\textrm{hill}}r_\parallel & |r_\parallel|\leq d/2,\\
        k_{\textrm{trap}}(|r_\parallel| - d/2)\textrm{sgn}(r_\parallel) & |r_\parallel| > d/2,
    \end{cases}
\end{eqnarray}
where $d = 3\mu$m is the trap separation, $k_\textrm{trap} = 0.1 \textrm{pN}/\textrm{nm}$ is the stiffness, and $k_\textrm{hill} = 8h/d^2$ with $h=8 \textrm{pN} \cdot \textrm{nm}$ being the height of the central hill. 
The traps are arranged vertically and horizontally in a 2D square lattice with lattice constant $a = 8.374 \mu$m as shown in Fig: \ref{fig:system}c).

The term $\mathbf{F}_i^{\textrm{dd}}$ represents the magnetic dipole-dipole interactions, described by the potential 
\begin{eqnarray}\label{eq:dipole_interactions}
    U_{ij} = -\dfrac{\mu_0 m^2}{4\pi r_{ij}^3}  \left[ 3\cos^2\phi - 1 \right],
\end{eqnarray}
where $\mathbf{m}$ is the magnetic moment of the colloids, given by $\mathbf{m} = V\chi\mathbf{B}/\mu_0$, and $\phi$ is the angle between the magnetic field vector, $\mathbf{B}$ and the vector joining both particles $\mathbf{r}_{ij} = \mathbf{r}_{i}-\mathbf{r}_{j}$.
\newchange{The angle $\theta$ is the polar angle between the magnetic field and the vertical direction, $\hat{z}$}
In the expression for the magnetic moment, $V$ is the colloid volume, and $\chi$ is the susceptibility. In our setup, colloidal particles have a radius of $r = 1.4 \mu$m and magnetic susceptibility $\chi = 0.4$, in agreement with values measured with commercially available particles \cite{martinez-pedrero_colloidal_2015}. \newchange{We neglect many body effects in our simulations, as particles are never close enough for these to be significant.}
All simulations were made with a timestep of $0.1$ms and a sample size of $n=30$ vertices per side, i.e. $1800$ particles. \newchange{These parameters were chosen to match what is experimentally available, except for the system size, the choice of periodic boundary conditions, and the variance of the central hill, which in our case is $\sigma_h=0$}.

\change{When the magnetic field is applied along one of the lattice basis vectors, it generates regions of attractive and repulsive interactions, separated by the so-called \emph{magic angle} $\phi = 54.7^\circ$, as illustrated in Fig.~\ref{fig:system}b). 
In a $4$-in or $4$-out vertex, all nearest-neighbor particles meet at an angle $\phi = 45\,\mathrm{deg} < \phi_c$, so their interactions are attractive. 
In consequence, particles in neighboring traps aligned with the magnetic field interact attractively, whereas particles in collinear traps perpendicular to the field experience repulsion. 
For traps oriented perpendicular to each other, the interaction can be either attractive or repulsive, depending on the relative positions of the particles. For example, as shown in Fig.~\ref{fig:system}b), when the particle in the horizontal trap is fixed close to the vertex, and the particle in the vertical trap is in the nearest stable point, they interact attractively, while when the vertical particle is in the opposite end of the trap, the interaction is attractive.

Although next-nearest-neighbor interactions across the vertex are mixed, they are weaker and won't overcome the energetic gain associated with bringing all particles close to the vertex. 
As a result, vertices with all particles pointing in or out are energetically favored, and we expect to observe predominantly charged vertices.
}

\section{\label{sec:results}Results and Discussion}

\begin{figure}[tbp]
    \includegraphics{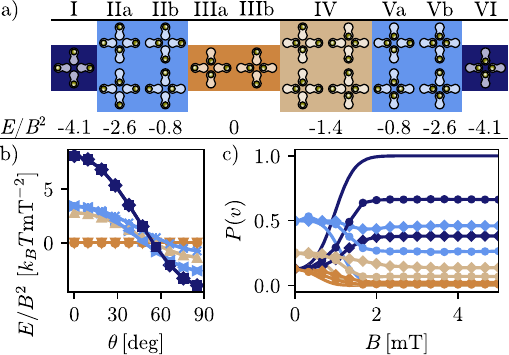}
    \caption{ \change{a) Color-coded  \change{vertex} types with their corresponding geometric vertex energy $E(v)/B^2$ with an in-plane field direction on the $x$ axis. b) Vertex energies with increasing tilting angle $\theta$ starting in the isotropic regime, i.e. orthogonal to the particle's plane. c) Vertices fraction with increasing magnetic field grouped by topological charge, for the mean-field model (solid, ---) line, fast simulations at $\dot{B} = 33.3\mu\mathrm{T}/\mathrm{s}$ (\textbf{-}$\blacklozenge$\textbf{-}), and slow simulations (\textbf{-}\textbullet\textbf{-}) at $\dot{B} = 0.333\mu\mathrm{T}/\mathrm{s}$. }}
    \label{fig:counts}
\end{figure}

With this relatively complicated interaction, we begin by estimating the energy hierarchy of all vertex types. This is \change{non-trivial} because the appearance of charged vertices (Types I, II, V, and VI, as defined in Fig.~\ref{fig:counts}), necessarily produce vertices with the opposite charge. The values shown in Fig. \ref{fig:counts} a) are calculated by a procedure similar to that of Refs. \cite{nisoli_unexpected_2018, oguz_topology_2020}. For each vertex, the \emph{naive} energy is given by the sum of all pair interactions between its components. The effective energy, $E_\mathrm{eff}$, is calculated as the average between the \emph{naive} energies of the vertex and its complement; that is, the vertex that appears from flipping all the particles. 

Using this procedure we reproduce the vertex hierarchy \change{known for a field orthogonal to the particle's plane} ($\theta = 0$), as shown in Fig. \ref{fig:counts} b). \change{Consequently, this model predicts that in our setup  ($\theta=90\deg$)}, the expected ground state is composed of highly charged vertices of types I and VI. 
Furthermore, we can plug the effective energy of vertex type $v$ into the Boltzmann relation, $P(v)\sim \exp(-E(v)/k_BT)$, to obtain the relative fraction of vertices of each type that are expected in a system in an equilibrium state. These expected vertex counts are shown as a solid line in Fig. \ref{fig:counts} c), where it can be seen that at a field of $B=3 \mathrm{mT}$, the equilibrium configuration should be composed purely of types I and VI. \change{For the square lattice, this ground state is twofold degenerate and can be realized by arranging $q = \pm 4$ topological charges antiferromagnetically in a checkerboard pattern, as illustrated in Fig.~\ref{fig:system}c).}

\change{Here we should note that this process of adjusting the energy hierarchy by modifying the system symmetry has been proposed in Refs. \cite{ostinato_tracking_2025, oguz_topology_2020}. Specifically, Ref. \cite{ostinato_tracking_2025} applies an external field that tilts the whole potential landscape and favors some vertices of type IV instead of type III. At high values of the external field, this results in a new ferromagnetic ground state, where all spins point towards one corner, saturating the magnetization. We note here that our approach cannot be described as an external field, since the changes in the potential landscape can't be reduced to a force that depends only on a point in space. Instead, what we are modifying is the symmetry of the pair interactions in a way that breaks rotational symmetry but preserves inversion symmetry.} 

\change{To realize the simulations, the system was initialized in a random configuration, and then the field was increased at a constant quench rate $\dot{B}$. Two quench protocols were realized at different rates: a \emph{fast} protocol at $\dot{B}=33.3\mu{}T/s$ and a \emph{slow} protocol at $\dot{B}=0.333\mu T/s$. Note that even in the case of the fast protocol, the time taken to reach $5\mathrm{mT}$ is around 2.5 minutes, while for slow quench it's more than 4 hours. After the quench both systems are allowed to relax for one hour at a constant field of $B_f = 5 mT$.}

Figure \ref{fig:counts} shows how both protocols deviate strongly from the model, even if they stop their evolution at around the same field value $B\approx3\mathrm{mT}$. The fast quench produces slightly more vertices of types II and V than of the expected ground state, I and V, while the slow quench slightly increases the fraction of ground state vertices, but the final fraction is still around $0.6$. The fact that slowing down the quench produces results closer to the model indicates that something in the dynamics prevents the system from reaching the ground state. It might be that there is a quench rate slow enough to allow the system to equilibrate; however, the modest improvement from 2.5 minutes to 4 hours suggests that the time required to achieve the ground state is likely impracticable. 

\begin{figure}[tbp]
    \includegraphics[width=0.95\linewidth]{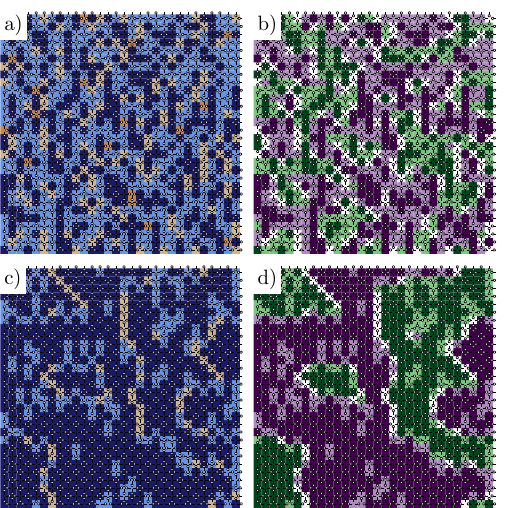}
    \caption{ Vertex Maps of the last frame of simulations in quenched anisitropic colloidal ice for fast (a,b) and slow (c,d) quenches. In panes a) and c), topological charges are directly colored corresponding to the colormap in Fig. \ref{fig:counts}. In b) and d) their color indicates to the value of their local \change{antiferromagnetic order parameter, with negative vertices shown as green, and positive vertices as purple. Regions with deep purple or green color indicate domains of incompatible ground state configuration.}}
    \label{fig:maps}
\end{figure}

We observe a typical spatial distribution of the final state in Fig. \ref{fig:maps}. In panels a) and c), we show the vertices according to the color scheme of Fig. \ref{fig:counts}. In this, the deep blue squares correspond to vertices of types I and VI and, therefore, to regions of local ground state configuration. All the pale blue and orange vertices are defects, with orange points being vertices of neutral charge. From these maps, we can conclude that the fast-quenched system is in a very disordered state, with domains of ground state vertices almost as large as defects. The slow quench is more ordered, with defects scattered on a field of ground state vertices. 

The domain structure becomes more evident when we plot a local order parameter in panels b) and d). Spatially, the ground state of the system is composed of a checkerboard pattern of vertices of types I and VI. We define the local order parameter
\begin{equation}
    \kappa_{x,y} = (-1)^{x+y}q_{xy},
\end{equation}
where $(x, y)$ are the components of the position vector in lattice units, and $q_{xy}$ is the charge of the vertex at position $(x,y)$. This order parameter acquires a value of $\pm1$ when the vertices are ordered according to the ground state and a value of $0$ for defects. As before, we can observe that the fast quench produces a nearly completely disordered structure, while the slow quench almost reaches a structure composed of two interpenetrated domains with scattered charged defects inside. The evolution of this order parameter for both quench rates can be seen in the videos S1 and S2. 

\begin{figure}[tb]
    \includegraphics[width=0.96\linewidth]{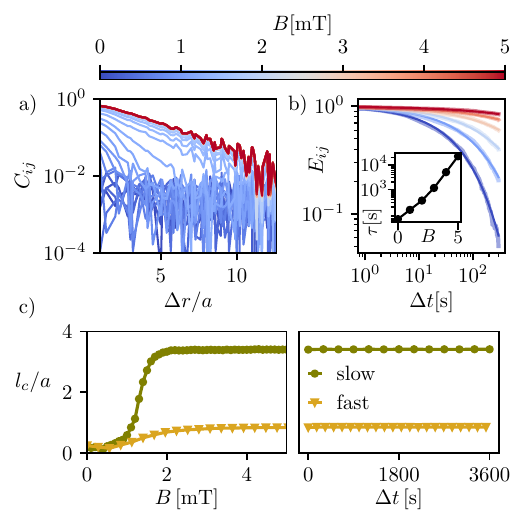}
    \caption{\label{fig:correlations}(a) Logarithmic plot of spin-spin spatial correlation functions for the slow quench and (b) Edwards-Anderson parameter at different magnetic fields. \newchange{The translucent lines are calculated from the simulation, and solid lines are a fit to a stretched exponential function $\exp\left[{-\left(\Delta t/\tau\right)^\beta}\right]$}. (c) Correlation length obtained \change{by fitting the spatial correlation functions with an exponential function, for both slow and fast simulations. The left panel shows how the correlation length remains steady at a constant field for an hour after the quench.}}
\end{figure}

To \change{characterize} the size of the domains we \change{compute} the spin-spin correlation function as follows:
\begin{align}
C(r_{ij}) = \left|\langle\sigma_i \cdot \sigma_j\rangle\right|.
\end{align}
In Fig. \ref{fig:correlations}a) we show the correlation functions for the slow quench simulations and see that these have generally decaying exponential functional forms, although for very low fields these correlations show almost only noise. We fit these functions with an exponential $\exp\left(-r_{ij}/l_c\right)$ to obtain the correlation length: that is, a measure of the average size of ordered regions in the lattice. By plotting $l_c$ in Fig. \ref{fig:correlations}c), we can observe how, in both slow and fast quenches, the ordered domains start growing but \change{saturate} for fields greater than $\sim 2 \mathrm{mT}$ \change{in agreement with the vertex counts from Fig: \ref{fig:counts} c)}. \change{The} correlation length can grow up to three lattice sites for slow quenches, but stays lower than one lattice site for fast quenches, indicating the system does evolve from its random configuration, but maintains a strongly disordered state. \newchange{This contrasts with the observations of Ref. \cite{libal_quenched_2020}, where the number of defects decreases due to the motion of the domain walls.} After the quench, the correlation length remains constant for up to an hour at constant field. This can also be seen in the second half of videos S1 and S2.

Figure \ref{fig:correlations}b) shows the Edwards-Anderson parameter for several values of the field. This data required a different protocol, where the field was increased in discrete steps. This allowed the system to evolve for some time in a given field so that the time correlation could be calculated as: 
\begin{align}
E (\Delta t) = \langle r_i(t)r_i(t+\Delta t) \rangle,
\end{align}
where the average is taken over $t$ and $i$. 
Surprisingly, particles still exhibit some mobility even at high fields, \change{where the vertex dynamics have effectively stopped evolving}. However, this reflects the local jumping of specific particles that have no consequence on the overall domain structure. 
\change{The slowing down of the relaxation appears to be similar to that observed experimentally in strongly confined microgel particles \cite{han_geometric_2008}. However, in that system, spin-lattice distortions appear from the quasi-2D displacements, which are not present in our system.} \newchange{The inset in Fig. 4b) shows the relaxation times $\tau$ as the field increases. To calculate it, we fit the time correlation with a stretched exponential function, $\exp\left[{-\left(\Delta t/\tau\right)^\beta}\right]$, shown in the figure as solid lines over the simulation data. The relaxation time clearly increases several orders of magnitude as the field increases, which indicates that the system's dynamics become arrested. The precise functional form, and the point at which $\tau$ diverges is left for future work.}

The spin freezing we observe arises due to thermal fluctuations at zero and low fields. Under these conditions, particles can jump between two equilibrium positions as the central hill barrier is in the order of $\sim 2k_B T$. As the field increases, the potential landscape experienced by each particle changes significantly, making transitions between the two positions a rare event, \change{even if the system's energetics favor one position considerably over the other.}

\begin{figure}[tb]
    \begin{center}
        \includegraphics{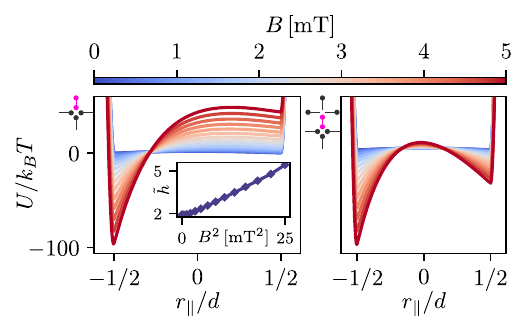}
    \end{center}
    \caption{Modified Potentials for a single vertex (left) and two coupled vertices (right) with increasing magnetic field when the highlighted particle is moved across the trap. The inset shows the trap's height $\tilde{h} = h/k_BT$ with squared field $B^2$. \change{The left panel shows the large increase in the potential barrier necessary to cross to the minimum energy configuration.}
    }\label{fig:energybarrier}
\end{figure}

To demonstrate this concept, Fig. \ref{fig:energybarrier} illustrates the modified potential of a vertical trap inside a single-vertex (left) and two coupled vertices (right). The potential is calculated along the direction of the trap highlighted in magenta. 
In both cases, as the field increases, the potential barrier between the two states increases, resulting in a decrease in the transition probability, even though one of the stable positions is more energetic than the other. The inset on the single-vertex shows the height of the potential barrier $\tilde{h} = h/k_BT$ measured at the center of the trap, which increases linearly with $B^2$. 
In the single-vertex case, we observe that the potential going from $+d/2\to-d/2$ consists of a long and shallow hill and a sudden drop as we approach $d/2$. This implies that, as previously computed, $q=+4$ vertices are more stable, but $q=+2$ vertices remain metastable due to the energetic barrier preventing transitions to the ground state. When a second vertex is added, the ground state keeps being a combination of charges $q=\pm4$, however, the metastable position $d/2$ is separated by a higher hill, making the $\pm2$ vertex pair more stable.

\section{Conclusions}

\change{In this paper, we studied the anisotropic regime in Artificial Colloidal Ice using both theory and simulations}. Our model suggest that when the magnetic field points in the direction of a lattice vector, the system has a ground state that locally maximizes the topological charge, thus forming an antiferromagnetic configuration of $\pm 4$ charges arranged in a checkerboard pattern. However, Brownian Dynamics simulations show that instead, the system \newchange{becomes arrested} in a metastable, disordered state, deviating significantly from the equilibrium theoretical predictions. Slower quenching rates improved agreement with the models, indicating that the intrinsic dynamics of the system prevented it from reaching equilibrium.

\newchange{The temporal correlations show that, as the field increases, the timescale for the spin flips diverges, and the system falls to a frozen state that stops evolving.
In disorder-free spin glass materials, it has been observed that freezing arises from small lattice distortions that induce enough disorder in the interactions for a glass state to appear, as observed in $\mathrm{Y}_2\mathrm{Mo}_2\mathrm{O}_7$ \cite{mitsumoto_spin-orbital_2020}.}
In our system such fluctuations are not present, because the traps are fixed in space, and instead the frozen state arises from magnetic interactions themselves; as the system is cooled, the potential barrier between the two equilibrium positions increases, eventually becoming large enough to prevent particles from thermally exploring the phase space, revealing a different type of spin freezing without quenched disorder. \newchange{This mechanism deviates significantly from the one observed in other colloidal models \cite{liangGlassTransitionMonolayers2025, zhouGlassySpinDynamics2017, han_geometric_2008} where the glassiness is induced by very short range interactions between anisotropic particles, or by the interaction with a  confinement, as well as from the mechanisms known to produce freezing in spin glass materials  \cite{mitsumoto_spin-orbital_2020, mitsumoto_replica_2023}.}
\newchange{Our system produces a dynamical freezing in a system with only one degree of freedom per particle. This reduced dimensionality allowed us to observe precisely the rise in the potential barrier that arrests the motion, while keeping the collective nature of the observed phenomena. We expect that our system can be used as model system for more complicated glassy systems.}

\change{Additionally,} we note that the behavior we observe here is markedly different from that shown in isotropic colloidal ice \cite{libal_quenched_2020}, where slower quenches also produced larger domains, but these domains continued to coarsen over time. The fact that our system freezes and stops evolving could make it suitable to observe the scalings predicted by the Kibble-Zurek mechanism \cite{del_campo_universality_2014} \newchange{by overcoming the difficulties observed in Ref. \cite{libal_quenched_2020}, where the coarsening of the domains after the transition meant that the characteristic lengthscale kept growing with time. We leave this research avenue for a future study.}

\acknowledgments{We thank Carolina Rodriguez-Gallo, Pietro Tierno, Yusheng Lei, and Xuan Feng for their fruitful discussions. LGAC acknowledges support from SECIHTI under its graduate scholarship program.}

\bibliography{bibliography.bib}

@article{mitsumoto_replica_2023,
	title = {Replica theory for disorder-free spin-lattice glass transition on a treelike simplex network},
	volume = {107},
	issn = {2469-9950, 2469-9969},
	url = {https://link.aps.org/doi/10.1103/PhysRevB.107.054412},
	doi = {10.1103/PhysRevB.107.054412},
	number = {5},
	urldate = {2025-11-27},
	journal = {Phys. Rev. B},
	author = {Mitsumoto, Kota and Yoshino, Hajime},
	month = feb,
	year = {2023},
	pages = {054412},
}

@article{ostinato_tracking_2025,
	title = {Tracking topological defects in a plasma state of a quenched colloidal ice},
	volume = {111},
	issn = {2469-9950, 2469-9969},
	url = {https://link.aps.org/doi/10.1103/PhysRevB.111.014414},
	doi = {10.1103/PhysRevB.111.014414},
	number = {1},
	urldate = {2025-11-27},
	journal = {Phys. Rev. B},
	author = {Ostinato, Mattia and Tierno, Pietro},
	month = jan,
	year = {2025},
	pages = {014414},
}

@article{zhang_unusual_2008,
	title = {Unusual {Slow} {Magnetic} {Relaxation} in {Helical} {Co}$_{\textrm{3}}$ ({OH})$_{\textrm{2}}$ {Ferrimagnetic} {Chain} {Based} {Cobalt} {Hydroxysulfates}},
	volume = {20},
	issn = {0897-4756, 1520-5002},
	url = {https://pubs.acs.org/doi/10.1021/cm7032542},
	doi = {10.1021/cm7032542},
	number = {6},
	urldate = {2025-10-02},
	journal = {Chem. Mater.},
	author = {Zhang, Xian-Ming and Li, Cui-Rui and Zhang, Xu-Hui and Zhang, Wei-Xiong and Chen, Xiao-Ming},
	month = mar,
	year = {2008},
	pages = {2298--2305},
}

@article{ortiz-ambriz_colloquium_2019,
	title = {\textit{{Colloquium}} : {Ice} rule and emergent frustration in particle ice and beyond},
	volume = {91},
	issn = {0034-6861, 1539-0756},
	shorttitle = {\textit{{Colloquium}}},
	url = {https://link.aps.org/doi/10.1103/RevModPhys.91.041003},
	doi = {10.1103/RevModPhys.91.041003},
	number = {4},
	urldate = {2020-10-12},
	journal = {Rev. Mod. Phys.},
	author = {Ortiz-Ambriz, Antonio and Nisoli, Cristiano and Reichhardt, Charles and Reichhardt, Cynthia J. O. and Tierno, Pietro},
	month = dec,
	year = {2019},
	pages = {041003},
}

@article{oguz_topology_2020,
	title = {Topology {Restricts} {Quasidegeneracy} in {Sheared} {Square} {Colloidal} {Ice}},
	volume = {124},
	copyright = {All rights reserved},
	issn = {0031-9007, 1079-7114},
	url = {https://link.aps.org/doi/10.1103/PhysRevLett.124.238003},
	doi = {10.1103/PhysRevLett.124.238003},
	number = {23},
	urldate = {2021-01-02},
	journal = {Phys. Rev. Lett.},
	author = {Oğuz, Erdal C. and Ortiz-Ambriz, Antonio and Shem-Tov, Hadas and Babià-Soler, Eric and Tierno, Pietro and Shokef, Yair},
	month = jun,
	year = {2020},
	pages = {238003},
}

@article{han_geometric_2008,
	title = {Geometric frustration in buckled colloidal monolayers},
	volume = {456},
	copyright = {http://www.springer.com/tdm},
	issn = {0028-0836, 1476-4687},
	url = {https://www.nature.com/articles/nature07595},
	doi = {10.1038/nature07595},
	number = {7224},
	urldate = {2025-03-20},
	journal = {Nature},
	author = {Han, Yilong and Shokef, Yair and Alsayed, Ahmed M. and Yunker, Peter and Lubensky, Tom C. and Yodh, Arjun G.},
	month = dec,
	year = {2008},
	pages = {898--903},
}

@article{schiffer_frustration_1995,
	title = {Frustration {Induced} {Spin} {Freezing} in a {Site}-{Ordered} {Magnet}: {Gadolinium} {Gallium} {Garnet}},
	volume = {74},
	copyright = {http://link.aps.org/licenses/aps-default-license},
	issn = {0031-9007, 1079-7114},
	shorttitle = {Frustration {Induced} {Spin} {Freezing} in a {Site}-{Ordered} {Magnet}},
	url = {https://link.aps.org/doi/10.1103/PhysRevLett.74.2379},
	doi = {10.1103/PhysRevLett.74.2379},
	number = {12},
	urldate = {2024-07-22},
	journal = {Phys. Rev. Lett.},
	author = {Schiffer, P. and Ramirez, A. P. and Huse, D. A. and Gammel, P. L. and Yaron, U. and Bishop, D. J. and Valentino, A. J.},
	month = mar,
	year = {1995},
	pages = {2379--2382},
}

@article{libal_quenched_2020,
	title = {Quenched {Dynamics} of {Artificial} {Spin} {Ice}: {Coarsening} versus {Kibble}-{Zurek}},
	volume = {2},
	issn = {2643-1564},
	shorttitle = {Quenched {Dynamics} of {Artificial} {Spin} {Ice}},
	url = {http://arxiv.org/abs/1908.05328},
	doi = {10.1103/PhysRevResearch.2.033433},
	number = {3},
	urldate = {2020-11-25},
	journal = {Phys. Rev. Research},
	author = {Libal, A. and del Campo, A. and Nisoli, C. and Reichhardt, C. and Reichhardt, C. J. O.},
	month = sep,
	year = {2020},
	note = {arXiv: 1908.05328},
	pages = {033433},
}

@article{nisoli_unexpected_2018,
	title = {Unexpected {Phenomenology} in {Particle}-{Based} {Ice} {Absent} in {Magnetic} {Spin} {Ice}},
	volume = {120},
	issn = {0031-9007, 1079-7114},
	url = {https://link.aps.org/doi/10.1103/PhysRevLett.120.167205},
	doi = {10.1103/PhysRevLett.120.167205},
	number = {16},
	urldate = {2020-11-18},
	journal = {Phys. Rev. Lett.},
	author = {Nisoli, Cristiano},
	month = apr,
	year = {2018},
	pages = {167205},
}

@article{wang_artificial_2006,
	title = {Artificial ‘spin ice’ in a geometrically frustrated lattice of nanoscale ferromagnetic islands},
	volume = {439},
	issn = {0028-0836, 1476-4687},
	url = {http://www.nature.com/articles/nature04447},
	doi = {10.1038/nature04447},
	number = {7074},
	urldate = {2020-10-13},
	journal = {Nature},
	author = {Wang, R. F. and Nisoli, C. and Freitas, R. S. and Li, J. and McConville, W. and Cooley, B. J. and Lund, M. S. and Samarth, N. and Leighton, C. and Crespi, V. H. and Schiffer, P.},
	month = jan,
	year = {2006},
	note = {tex.ids: wang\_artificial\_2006},
	pages = {303--306},
}

@article{ortiz-ambriz_engineering_2016,
	title = {Engineering of frustration in colloidal artificial ices realized on microfeatured grooved lattices},
	volume = {7},
	issn = {2041-1723},
	url = {http://www.nature.com/articles/ncomms10575},
	doi = {10.1038/ncomms10575},
	number = {1},
	urldate = {2020-10-13},
	journal = {Nat. Commun.},
	author = {Ortiz-Ambriz, Antonio and Tierno, Pietro},
	month = apr,
	year = {2016},
	pages = {10575},
}

@article{martinez-pedrero_colloidal_2015,
	title = {Colloidal {Microworms} {Propelling} via a {Cooperative} {Hydrodynamic} {Conveyor} {Belt}},
	volume = {115},
	copyright = {All rights reserved},
	issn = {0031-9007, 1079-7114},
	url = {https://link.aps.org/doi/10.1103/PhysRevLett.115.138301},
	doi = {10.1103/PhysRevLett.115.138301},
	number = {13},
	urldate = {2021-01-02},
	journal = {Phys. Rev. Lett.},
	author = {Martinez-Pedrero, Fernando and Ortiz-Ambriz, Antonio and Pagonabarraga, Ignacio and Tierno, Pietro},
	month = sep,
	year = {2015},
	pages = {138301},
}

@article{mengotti_real-space_2011,
	title = {Real-space observation of emergent magnetic monopoles and associated {Dirac} strings in artificial kagome spin ice},
	volume = {7},
	issn = {1745-2473, 1745-2481},
	url = {http://www.nature.com/articles/nphys1794},
	doi = {10.1038/nphys1794},
	number = {1},
	urldate = {2020-10-13},
	journal = {Nature Phys},
	author = {Mengotti, Elena and Heyderman, Laura J. and Rodríguez, Arantxa Fraile and Nolting, Frithjof and Hügli, Remo V. and Braun, Hans-Benjamin},
	month = jan,
	year = {2011},
	note = {tex.ids: mengotti\_realspace\_2011a},
	pages = {68--74},
}

@article{morley_vogel-fulcher-tammann_2017,
	title = {Vogel-{Fulcher}-{Tammann} freezing of a thermally fluctuating artificial spin ice probed by x-ray photon correlation spectroscopy},
	volume = {95},
	issn = {2469-9950, 2469-9969},
	url = {https://link.aps.org/doi/10.1103/PhysRevB.95.104422},
	doi = {10.1103/PhysRevB.95.104422},
	number = {10},
	urldate = {2023-09-22},
	journal = {Phys. Rev. B},
	author = {Morley, S. A. and Alba Venero, D. and Porro, J. M. and Riley, S. T. and Stein, A. and Steadman, P. and Stamps, R. L. and Langridge, S. and Marrows, C. H.},
	month = mar,
	year = {2017},
	pages = {104422},
}

@article{sullow_spin_1997,
	title = {Spin {Glass} {Behavior} in {URh} 2 {Ge} 2},
	volume = {78},
	copyright = {http://link.aps.org/licenses/aps-default-license},
	issn = {0031-9007, 1079-7114},
	url = {https://link.aps.org/doi/10.1103/PhysRevLett.78.354},
	doi = {10.1103/PhysRevLett.78.354},
	number = {2},
	urldate = {2024-10-11},
	journal = {Phys. Rev. Lett.},
	author = {Süllow, S. and Nieuwenhuys, G. J. and Menovsky, A. A. and Mydosh, J. A. and Mentink, S. A. M. and Mason, T. E. and Buyers, W. J. L.},
	month = jan,
	year = {1997},
	pages = {354--357},
}

@article{snyder_low-temperature_2004,
	title = {Low-temperature spin freezing in the {Dy} 2 {Ti} 2 {O} 7 spin ice},
	volume = {69},
	copyright = {http://link.aps.org/licenses/aps-default-license},
	issn = {1098-0121, 1550-235X},
	url = {https://link.aps.org/doi/10.1103/PhysRevB.69.064414},
	doi = {10.1103/PhysRevB.69.064414},
	number = {6},
	urldate = {2024-10-11},
	journal = {Phys. Rev. B},
	author = {Snyder, J. and Ueland, B. G. and Slusky, J. S. and Karunadasa, H. and Cava, R. J. and Schiffer, P.},
	month = feb,
	year = {2004},
	pages = {064414},
}

@article{snyder_how_2001,
	title = {How ‘spin ice’ freezes},
	volume = {413},
	copyright = {http://www.springer.com/tdm},
	issn = {0028-0836, 1476-4687},
	url = {https://www.nature.com/articles/35092516},
	doi = {10.1038/35092516},
	number = {6851},
	urldate = {2024-10-11},
	journal = {Nature},
	author = {Snyder, J. and Slusky, J. S. and Cava, R. J. and Schiffer, P.},
	month = sep,
	year = {2001},
	pages = {48--51},
}

@article{reimers_short-range_1991,
	title = {Short-range magnetic ordering in the highly frustrated pyrochlore {Y} 2 {Mn} 2 {O} 7},
	volume = {43},
	copyright = {http://link.aps.org/licenses/aps-default-license},
	issn = {0163-1829, 1095-3795},
	url = {https://link.aps.org/doi/10.1103/PhysRevB.43.3387},
	doi = {10.1103/PhysRevB.43.3387},
	number = {4},
	urldate = {2024-10-11},
	journal = {Phys. Rev. B},
	author = {Reimers, J. N. and Greedan, J. E. and Kremer, R. K. and Gmelin, E. and Subramanian, M. A.},
	month = feb,
	year = {1991},
	pages = {3387--3394},
}

@article{raju_magnetic-susceptibility_1992,
	title = {Magnetic-susceptibility and specific-heat studies of spin-glass-like ordering in the pyrochlore compounds {R} 2 {Mo} 2 {O} 7 ( \textit{{R}} ={Y}, {Sm}, or {Gd})},
	volume = {46},
	copyright = {http://link.aps.org/licenses/aps-default-license},
	issn = {0163-1829, 1095-3795},
	url = {https://link.aps.org/doi/10.1103/PhysRevB.46.5405},
	doi = {10.1103/PhysRevB.46.5405},
	number = {9},
	urldate = {2024-10-11},
	journal = {Phys. Rev. B},
	author = {Raju, N. P. and Gmelin, E. and Kremer, R. K.},
	month = sep,
	year = {1992},
	pages = {5405--5411},
}

@article{mitsumoto_spin-orbital_2020,
	title = {Spin-{Orbital} {Glass} {Transition} in a {Model} of a {Frustrated} {Pyrochlore} {Magnet} without {Quenched} {Disorder}},
	volume = {124},
	issn = {0031-9007, 1079-7114},
	url = {https://link.aps.org/doi/10.1103/PhysRevLett.124.087201},
	doi = {10.1103/PhysRevLett.124.087201},
	number = {8},
	urldate = {2024-10-11},
	journal = {Phys. Rev. Lett.},
	author = {Mitsumoto, Kota and Hotta, Chisa and Yoshino, Hajime},
	month = feb,
	year = {2020},
	pages = {087201},
}

@article{gaulin_spin_1992,
	title = {Spin freezing in the geometrically frustrated pyrochlore antiferromagnet {Tb} 2 {Mo} 2 {O} 7},
	volume = {69},
	copyright = {http://link.aps.org/licenses/aps-default-license},
	issn = {0031-9007},
	url = {https://link.aps.org/doi/10.1103/PhysRevLett.69.3244},
	doi = {10.1103/PhysRevLett.69.3244},
	number = {22},
	urldate = {2024-10-11},
	journal = {Phys. Rev. Lett.},
	author = {Gaulin, B. D. and Reimers, J. N. and Mason, T. E. and Greedan, J. E. and Tun, Z.},
	month = nov,
	year = {1992},
	pages = {3244--3247},
}

@article{goremychkin_spin-glass_2008,
	title = {Spin-glass order induced by dynamic frustration},
	volume = {4},
	copyright = {http://www.springer.com/tdm},
	issn = {1745-2473, 1745-2481},
	url = {https://www.nature.com/articles/nphys1028},
	doi = {10.1038/nphys1028},
	number = {10},
	urldate = {2024-10-11},
	journal = {Nature Phys},
	author = {Goremychkin, E. A. and Osborn, R. and Rainford, B. D. and Macaluso, R. T. and Adroja, D. T. and Koza, M.},
	month = oct,
	year = {2008},
	pages = {766--770},
}

@article{edwards_theory_1975,
	title = {Theory of spin glasses},
	volume = {5},
	issn = {0305-4608},
	url = {https://iopscience.iop.org/article/10.1088/0305-4608/5/5/017},
	doi = {10.1088/0305-4608/5/5/017},
	number = {5},
	urldate = {2024-10-11},
	journal = {J. Phys. F: Met. Phys.},
	author = {Edwards, S F and Anderson, P W},
	month = may,
	year = {1975},
	pages = {965--974},
}

@article{saccone_real-space_2023,
	title = {Real-space observation of ergodicity transitions in artificial spin ice},
	volume = {14},
	issn = {2041-1723},
	url = {https://www.nature.com/articles/s41467-023-41235-4},
	doi = {10.1038/s41467-023-41235-4},
	number = {1},
	urldate = {2024-07-26},
	journal = {Nat Commun},
	author = {Saccone, Michael and Caravelli, Francesco and Hofhuis, Kevin and Dhuey, Scott and Scholl, Andreas and Nisoli, Cristiano and Farhan, Alan},
	month = sep,
	year = {2023},
	pages = {5674},
}

@article{rodriguez-gallo_geometrical_2023,
	title = {Geometrical control of topological charge transfer in {Shakti}-{Cairo} colloidal ice},
	volume = {6},
	copyright = {All rights reserved},
	issn = {2399-3650},
	url = {https://www.nature.com/articles/s42005-023-01236-7},
	doi = {10.1038/s42005-023-01236-7},
	number = {1},
	urldate = {2024-01-10},
	journal = {Commun Phys},
	author = {Rodríguez-Gallo, Carolina and Ortiz-Ambriz, Antonio and Nisoli, Cristiano and Tierno, Pietro},
	month = may,
	year = {2023},
	pages = {113},
}

@article{del_campo_universality_2014,
	title = {Universality of phase transition dynamics: {Topological} defects from symmetry breaking},
	volume = {29},
	issn = {0217-751X, 1793-656X},
	shorttitle = {Universality of phase transition dynamics},
	url = {https://www.worldscientific.com/doi/abs/10.1142/S0217751X1430018X},
	doi = {10.1142/S0217751X1430018X},
	number = {08},
	urldate = {2025-05-20},
	journal = {Int. J. Mod. Phys. A},
	author = {Del Campo, Adolfo and Zurek, Wojciech H.},
	month = mar,
	year = {2014},
	pages = {1430018},
}

@article{andersson_thermally_2016,
	title = {Thermally induced magnetic relaxation in square artificial spin ice},
	volume = {6},
	issn = {2045-2322},
	url = {https://www.nature.com/articles/srep37097},
	doi = {10.1038/srep37097},
	number = {1},
	urldate = {2025-05-12},
	journal = {Sci Rep},
	author = {Andersson, M. S. and Pappas, S. D. and Stopfel, H. and Östman, E. and Stein, A. and Nordblad, P. and Mathieu, R. and Hjörvarsson, B. and Kapaklis, V.},
	month = nov,
	year = {2016},
	pages = {37097},
}

@article{zimmerman_low-temperature_1960,
	title = {Low-temperature specific heat of dilute {Cu}-{Mn} alloys},
	volume = {17},
	copyright = {https://www.elsevier.com/tdm/userlicense/1.0/},
	issn = {00223697},
	url = {https://linkinghub.elsevier.com/retrieve/pii/0022369760901748},
	doi = {10.1016/0022-3697(60)90174-8},
	number = {1-2},
	urldate = {2025-05-12},
	journal = {Journal of Physics and Chemistry of Solids},
	author = {Zimmerman, J.E. and Hoare, F.E.},
	month = dec,
	year = {1960},
	pages = {52--56},
}

@article{rougemaille_artificial_2011,
	title = {Artificial {Kagome} {Arrays} of {Nanomagnets}: {A} {Frozen} {Dipolar} {Spin} {Ice}},
	volume = {106},
	copyright = {http://link.aps.org/licenses/aps-default-license},
	issn = {0031-9007, 1079-7114},
	shorttitle = {Artificial {Kagome} {Arrays} of {Nanomagnets}},
	url = {https://link.aps.org/doi/10.1103/PhysRevLett.106.057209},
	doi = {10.1103/PhysRevLett.106.057209},
	number = {5},
	urldate = {2025-05-12},
	journal = {Phys. Rev. Lett.},
	author = {Rougemaille, N. and Montaigne, F. and Canals, B. and Duluard, A. and Lacour, D. and Hehn, M. and Belkhou, R. and Fruchart, O. and El Moussaoui, S. and Bendounan, A. and Maccherozzi, F.},
	month = feb,
	year = {2011},
	pages = {057209},
}

@article{libal_realizing_2006,
	title = {Realizing {Colloidal} {Artificial} {Ice} on {Arrays} of {Optical} {Traps}},
	volume = {97},
	copyright = {http://link.aps.org/licenses/aps-default-license},
	issn = {0031-9007, 1079-7114},
	url = {https://link.aps.org/doi/10.1103/PhysRevLett.97.228302},
	doi = {10.1103/PhysRevLett.97.228302},
	number = {22},
	urldate = {2025-05-12},
	journal = {Phys. Rev. Lett.},
	author = {Libál, A. and Reichhardt, C. and Reichhardt, C. J. Olson},
	month = nov,
	year = {2006},
	pages = {228302},
}

@article{lao_classical_2018,
	title = {Classical topological order in the kinetics of artificial spin ice},
	volume = {14},
	issn = {1745-2473, 1745-2481},
	url = {https://www.nature.com/articles/s41567-018-0077-0},
	doi = {10.1038/s41567-018-0077-0},
	number = {7},
	urldate = {2025-05-12},
	journal = {Nature Phys},
	author = {Lao, Yuyang and Caravelli, Francesco and Sheikh, Mohammed and Sklenar, Joseph and Gardeazabal, Daniel and Watts, Justin D. and Albrecht, Alan M. and Scholl, Andreas and Dahmen, Karin and Nisoli, Cristiano and Schiffer, Peter},
	month = jul,
	year = {2018},
	pages = {723--727},
}

@article{krimmel_spin-glass_1999,
	title = {Spin-glass behavior in {PrAu} 2 {Si} 2},
	volume = {59},
	copyright = {http://link.aps.org/licenses/aps-default-license},
	issn = {0163-1829, 1095-3795},
	url = {https://link.aps.org/doi/10.1103/PhysRevB.59.R6604},
	doi = {10.1103/PhysRevB.59.R6604},
	number = {10},
	urldate = {2025-05-12},
	journal = {Phys. Rev. B},
	author = {Krimmel, A. and Hemberger, J. and Nicklas, M. and Knebel, G. and Trinkl, W. and Brando, M. and Fritsch, V. and Loidl, A. and Ressouche, E.},
	month = mar,
	year = {1999},
	pages = {R6604--R6607},
}

@article{hadouchi_unconventional_2019,
	title = {Unconventional spin-glass-like state in {AgCo2V3O10}, the novel magnetically frustrated material},
	volume = {491},
	issn = {03048853},
	url = {https://linkinghub.elsevier.com/retrieve/pii/S030488531931296X},
	doi = {10.1016/j.jmmm.2019.165623},
	urldate = {2025-05-12},
	journal = {Journal of Magnetism and Magnetic Materials},
	author = {Hadouchi, Mohammed and Assani, Abderrazzak and Saadi, Mohamed and Kopelevich, Yakov and Da Silva, Robson R. and Lahmar, Abdelilah and Bouyanfif, Houssny and El Marssi, Mimoun and El Ammari, Lahcen},
	month = dec,
	year = {2019},
	pages = {165623},
}

@article{de_nobel_specific_1959,
	title = {Specific heats of dilute alloys of manganese in silver and copper at low temperatures and in magnetic fields},
	volume = {25},
	copyright = {https://www.elsevier.com/tdm/userlicense/1.0/},
	issn = {00318914},
	url = {https://linkinghub.elsevier.com/retrieve/pii/0031891459900187},
	doi = {10.1016/0031-8914(59)90018-7},
	number = {7-12},
	urldate = {2025-05-12},
	journal = {Physica},
	author = {De Nobel, J. and Du Chatenier, F.J.},
	month = jan,
	year = {1959},
	pages = {969--979},
}

@article{saccone_direct_2022,
	title = {Direct observation of a dynamical glass transition in a nanomagnetic artificial {Hopfield} network},
	volume = {18},
	issn = {1745-2473, 1745-2481},
	url = {https://www.nature.com/articles/s41567-022-01538-7},
	doi = {10.1038/s41567-022-01538-7},
	number = {5},
	urldate = {2024-10-16},
	journal = {Nat. Phys.},
	author = {Saccone, Michael and Caravelli, Francesco and Hofhuis, Kevin and Parchenko, Sergii and Birkhölzer, Yorick A. and Dhuey, Scott and Kleibert, Armin and Van Dijken, Sebastiaan and Nisoli, Cristiano and Farhan, Alan},
	month = may,
	year = {2022},
	pages = {517--521},
}

@article{thygesen_orbital_2017,
	title = {Orbital {Dimer} {Model} for the {Spin}-{Glass} {State} in {Y} 2 {Mo} 2 {O} 7},
	volume = {118},
	copyright = {http://link.aps.org/licenses/aps-default-license},
	issn = {0031-9007, 1079-7114},
	url = {https://link.aps.org/doi/10.1103/PhysRevLett.118.067201},
	doi = {10.1103/PhysRevLett.118.067201},
	number = {6},
	urldate = {2024-10-11},
	journal = {Phys. Rev. Lett.},
	author = {Thygesen, Peter M. M. and Paddison, Joseph A. M. and Zhang, Ronghuan and Beyer, Kevin A. and Chapman, Karena W. and Playford, Helen Y. and Tucker, Matthew G. and Keen, David A. and Hayward, Michael A. and Goodwin, Andrew L.},
	month = feb,
	year = {2017},
	pages = {067201},
}

@article{liangGlassTransitionMonolayers2025,
  title = {Glass {{Transition}} in {{Monolayers}} of {{Rough Colloidal Ellipsoids}}},
  author = {Liang, Jian and Feng, Xuan and Zheng, Ning and Wang, Huaguang and Ni, Ran and Zhang, Zexin},
  year = 2025,
  month = jan,
  journal = {Physical Review Letters},
  volume = {134},
  number = {3},
  pages = {038202},
  publisher = {American Physical Society},
  doi = {10.1103/PhysRevLett.134.038202},
  urldate = {2025-05-24}
}

@article{zhouGlassySpinDynamics2017,
  title = {Glassy {{Spin Dynamics}} in {{Geometrically Frustrated Buckled Colloidal Crystals}}},
  author = {Zhou, Di and Wang, Feng and Li, Bo and Lou, Xiaojie and Han, Yilong},
  year = 2017,
  month = may,
  journal = {Physical Review X},
  volume = {7},
  number = {2},
  pages = {021030},
  publisher = {American Physical Society},
  doi = {10.1103/PhysRevX.7.021030},
  urldate = {2023-09-13}
}

\end{document}